\documentclass[reprint,amsmath,amssymb,aps,pre,nofootinbib]{revtex4-2}

\usepackage{}
\usepackage{}
\usepackage{color}
\usepackage{graphicx}
\usepackage{hyperref}
\usepackage{subfigure}

\newcommand{\B}[1]{\mbox{\boldmath${#1}$\unboldmath}}

\begin{document}

\title{2D mobile breather scattering in a hexagonal crystal lattice}

\author{J\={a}nis Baj\={a}rs} \affiliation{Faculty of Physics, Mathematics and
  Optometry, University of Latvia, Jelgavas Street 3, Riga, LV-1004,
  Latvia} \email{Janis.Bajars@lu.lv}

\author{J. Chris Eilbeck} 
\affiliation{Maxwell Institute and School of
  Mathematical \& Computer Sciences, Heriot-Watt University,
  Edinburgh, EH14 4AS, United Kingdom} 
\email{J.C.Eilbeck@hw.ac.uk}

\author{Benedict Leimkuhler}
\affiliation{Maxwell Institute and School of Mathematics, The
  University of Edinburgh, James Clerk Maxwell Building, The King's
  Buildings, Mayfield Road, Edinburgh, EH9 3JZ, United Kingdom}
\email{B.Leimkuhler@ed.ac.uk}
\date{\today}

\begin{abstract}
  We describe, for the first time, the full 2D scattering of
  long-lived breathers in a model hexagonal lattice of atoms.  The
  chosen system, representing an idealized model of mica, combines a
  Lennard-Jones interatomic potential with an ``egg-box" harmonic
  potential well surface.  We investigate the dependence of breather
  properties on the ratio of the well depths associated to the
  interaction and on-site potentials. High values of this ratio lead
  to large spatial displacements in adjacent chains of atoms and thus
  enhance the two dimensional character of the quasi-one-dimensional
  breather solutions. This effect is further investigated during
  breather-breather collisions by following the constrained energy
  density function in time for a set of randomly excited mobile
  breather solutions.  Certain collisions lead to 60$^\circ$
  scattering, and collisions of mobile and stationary breathers can
  generate a rich variety of states.
\end{abstract}

\maketitle

The nature of mysterious particle-like tracks in muscovite mica
crystals have attracted much recent interest since
  Russell's first observations over 50 years ago \cite{Ru67}.  Russell
  suggested that some of them were caused by localized vibrational
  modes (which he called quodons) in the K-K layer of mica
  \cite{Ru91,ram20}.  This hypothesis has lead to a number of simulations
of breathers in model hexagonal lattices with on-site potentials
\cite{MaEiRu98,MaEiRu00,bel14a,BaEiLe15}.  The surprising conclusion
of these studies is that in 2D, localized single breathers can travel along the main crystal directions of the lattice with little
attenuation or lateral spreading.

In this note we move beyond the case of single breathers by examining
breather-breather collisions.  We present evidence that breathers are
remarkably robust to collisions, and scattering through some multiple
of 60$^\circ$ into another crystal direction is frequently observed in
some circumstances. In addition we examine ensembles of initial
conditions for breather-breather collisions to begin to understand how
the relative angles and phases of the breathers affect their
interactions.

Our simplified 2D model of the hexagonal K-K sheet layer in mica
crystal \cite{BaEiLe15} is based on the following dimensionless
Hamiltonian which describes the classical dynamics of $N$ potassium
atoms:
\begin{align}\label{Hamilt}
  H &= K_E + U + V_{c}  \\ \nonumber
  &= \sum_{n=1}^{N} \biggl( \frac{1}{2} |\dot{\B{r}}_{n}|^2 +
    U(\B{r}_{n}) + \frac{1}{2} \sum_{\genfrac{}{}{0pt}{2}{n'=1}{n'\neq n}}^{N}
    V_{c}(|\B{r}_{n}-\B{r}_{n'}|) \biggr),
\end{align}
where $K_E$ is the kinetic energy, $U$ is the on-site potential energy
(modelling forces from atoms above and below the K-K sheet),
$V_{c}$ is the radial interparticle potential of the potassium atoms
with a cut-off radius $r_c$, $\B{r}_n\in\mathbb{R}^2$ is the 2D
position vector of the $n^{th}$ K atom, $\dot{\B{r}}_{n}$ is its time
derivative, and $|\cdot|$ is the Euclidean distance.  
Note that no motion in the $z$-direction is allowed.  Any mention of
``transverse'' in the following means in-plane motion transverse to
the breather propagation line.

The dimensionless  on-site potential $U$ is modelled as a smooth
periodic function resembling an egg-box carton with hexagonal symmetry:
\begin{align}\label{OnSiteFunc}
  U(x,y) = &\tfrac{2}{3} \left(1-\tfrac{1}{3}  
             \left( \cos{ \biggl( \tfrac{2\pi(\sqrt{3}x-y)}
             {\sqrt{3}}\biggr)} \right.\right.\\ \nonumber
           & \qquad + \left.\left.\cos{
             \biggl( \tfrac{2\pi(\sqrt{3}x+y)}{\sqrt{3}} \biggr) }
             + \cos{\biggl(
             \tfrac{4\pi{y}}{\sqrt{3}}}
             \right) \right),
\end{align}
where $x$ and $y$ are configurational coordinates, $\B{r}_n=(x,y)$.
Importantly, in any of the three crystallographic lattice directions
with direction cosine vectors $(1,0)^{T}$ and
$(1/2,\pm\sqrt{3}/2)^{T}$, the on-site potential (\ref{OnSiteFunc}) is
a cosine, so the model reduces to a special case of the
Frenkel–-Kontorova model.  The 1D atomic chains in the $(1,0)^{T}$
lattice direction are denoted by $y_{m}$, where $m \in\mathbb{Z}$.
The interatomic interactions of K atoms are modelled by a scaled
Lennard-Jones potential $V_{LJ}(r)$ with cut-off radius $r_{c}$, i.e.
\begin{equation}
\label{CutPot}
V_{c}(r) = \epsilon \left( \left( \frac{1}{r} \right)^{12} - 2 \left(
    \frac{1}{r} \right)^{6} \right) + \epsilon \sum_{j=0}^{4} A_{j}
\left( \frac{r}{r_{c}} \right)^{2j} ,
\end{equation}
if $0 < r \leq r_{c}$, and zero elsewhere.  The cut-off dimensionless
coefficients $A_{j}\to{0}$ when $r_{c}\to{\infty}$ are determined from
matching and continuity conditions on $V_{LJ}$ at well depth $r=1$ and
the cutoff $r_c$, respectively, see \cite{BaEiLe15} for more details.

In this paper we consider $r_{c}=3$ in dimensionless units. In
general, we did not observe qualitatively different results for
$r_{c}\ge\sqrt{3}$. This can be attributed to the rapid decay of the
Lennard-Jones potential (\ref{CutPot}) for $r\to\infty$.

The dimensionless parameter $\epsilon$ controls the relative strength
of the two potentials. If $\epsilon=0$ then the system decouples into
nonlinear oscillators whereas, for $\epsilon\to\infty$, the system
behaves as a Lennard-Jones fluid. Previous studies \cite{MaEiRu98}
suggest that propagating breather solutions are 
observed when the two potentials are of roughly equal relative
strength. This occurs at about $\epsilon = 0.05$ \cite{BaEiLe15}. The
same paper describes how mobile breather solutions for (\ref{Hamilt})
can be observed in the range $\epsilon\in[0.001,1]$.

We mention here that other studies of hexagonal lattices have mostly
concentrated on Morse lattices {\em without} an on-site potential, see
for example \cite{CEV16,SKBSCD20} and references
therein. Ref.~\cite{SKBSCD20} discusses general cases of crowdions,
also known as kinks, pulses which assymptote to different values in
different directions. Ref.~\cite{CEV16} discusses solitons collisions
with 2D scattering but the scattered pulses are not
long-lived. Breather collisions are discussed in \cite{KSCV14}, but no
cases involving $60^\circ$ are described.

To excite mobile discrete breathers, the simplest method is to
consider atoms in their dynamical equilibrium state, i.e.\ at the
bottom of each well of the on-site potential (\ref{OnSiteFunc}), and
excite three co-linear neighbouring atomic momenta in any
crystallographic lattice direction with the pattern
\begin{equation}\label{Pattern}
 \B{v}_{0} = \gamma (-1,2,-1)^{T},
\end{equation}
where the values of $\gamma\in\mathbb{R}$ depend on the choice of
$\epsilon$.  In contrast to other initial excitations such as single
kicks or more complex patterns, we find that this pattern produces
clean initial conditions for the excitation of mobile discrete
breather solutions, i.e.~produced very few phonons. Note that
although this initial pattern appears to sum to equal and opposite
kicks in each direction, the effect is not symmetric due to the
nonlinearities involved.

Similarly, by considering patterns involving four co-linear atomic
momenta $ \B{w}_{0} = \gamma (-1,2,-2,1)^{T} $, we are able to excite
stationary breathers.  In the present study we concentrate on
breather-breather interactions and therefore avoid the complications
that a higher phonon density would bring.

We integrate the Hamiltonian dynamics of the system with the second
order time reversible symplectic Verlet method \cite{AlTi89,LeRe05}.
In the following, all numerical examples are performed with time step
$\tau=0.04$ and periodic boundary conditions for different values of
$\epsilon$ and $\gamma$. We can define an energy density function at
each lattice point by assigning to each atom its kinetic energy and
on-site potential values as well as half of the interaction potential
values. For conciseness we refer to this as the ``lattice point
energy".  To obtain positive values we redefine
$H := H + |\min{\{H\}}|$ such that $H\ge{0}$. In all energy density
plots we interpolate $H$ over a square uniform mesh.

The initial excitation (\ref{Pattern}) leads \cite{BaEiLe15} to highly
localized mobile breather solutions propagating on a chain of atoms in
a crystallographic lattice direction with large displacements in the
$x$ direction, almost zero displacements in the $y$-axis direction and
with small displacements in both axis directions on the chains of
atoms adjacent to the main chain of atoms. In addition, the observed
mobile breathers are optical with internal\footnote{By ``internal'' we
  mean the carrier wave inside the slowly varying envelope of the
  breather.} frequencies above the phonon linear spectrum.

In considering breather collisions, there are three possibilities. The
first is inline or head-to-head collisions with two breathers on the
same chain but travelling in opposite directions.  These were first
looked at briefly in \cite{MaEiRu00}.  The second occurs when two
breathers approach each other along the same lattice vectors but on
adjacent parallel chains. The third occurs when two breathers approach
along different lattice vectors, i.e.~at an angle of a multiple of
$60^\circ$.

Before proceeding to the study of collisions, we investigate mobile breather lattice point energies and displacement properties depending on the values of the parameters $\epsilon$ and
$\gamma$. We consider a periodic rectangular lattice with hexagonal
symmetry of $N_{x}=100$ atoms in the $x$-axis direction and $N_{y}=16$
atoms in the $y$-axis direction placed in a mechanical equilibrium
state. We excite atoms by (\ref{Pattern}) in the $(1,0)^T$
crystallographic lattice direction in the middle of the lattice with
respect to the $y$-axis and integrate in time until the breather has
passed $1000$ lattice sites, that is, crossed the domain along the
$x$-axis direction $100$ times. The final computational time varies
depending on the parameter $\epsilon$ and $\gamma$ values. In what
follows, the main horizontal chain of atoms where the breather
propagates is indicated by $y_{m}$.

\begin{figure}
  \centering \subfigure[]{\label{fig:EpsA}
    \includegraphics[trim=0.1cm 0.1cm 0.8cm 0.5cm,clip=true,width=0.315\textwidth]{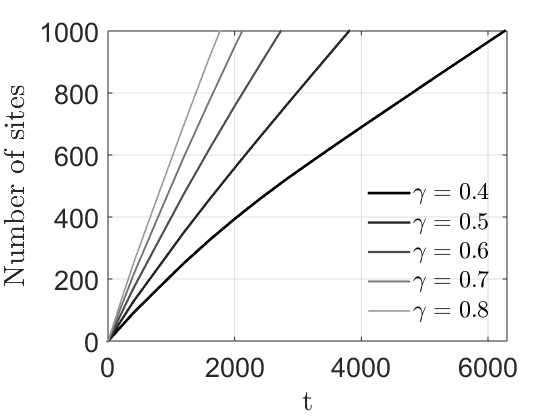}}
  \subfigure[]{\label{fig:EpsB}
    \includegraphics[trim=0.1cm 0.1cm 0.8cm 0.5cm,clip=true,width=0.315\textwidth]{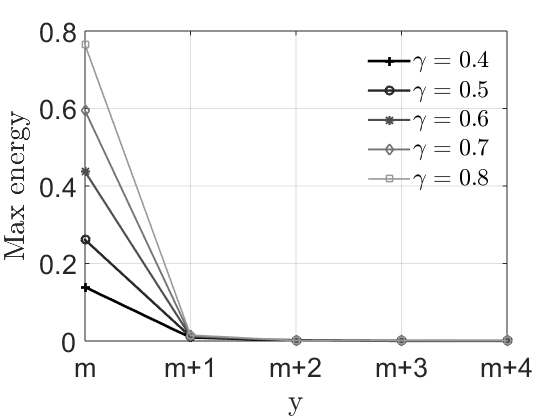}}
  \subfigure[]{\label{fig:EpsC}
    \includegraphics[trim=0.1cm 0.1cm 0.8cm 0.5cm,clip=true,width=0.315\textwidth]{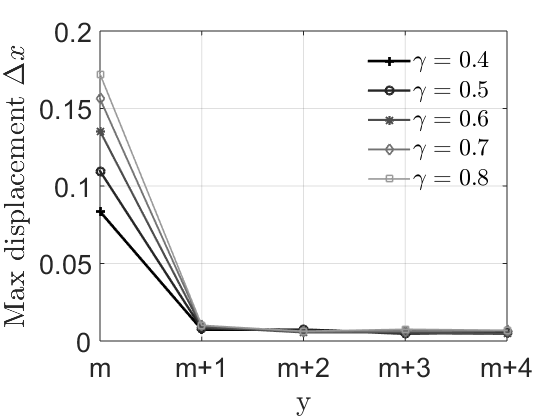}}
  \subfigure[]{\label{fig:EpsD}
    \includegraphics[trim=0.1cm 0.1cm 0.8cm 0.5cm,clip=true,width=0.315\textwidth]{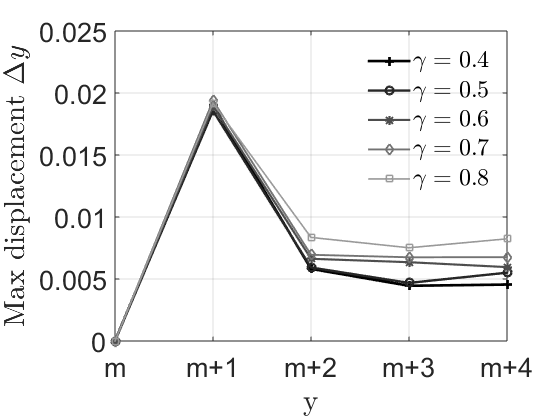}}
  \caption{Mobile breather properties for fixed dimensionless
    parameter value $\epsilon=0.05$ and different values of the excitation
    parameter $\gamma$.  (a) breather travel distance versus time in
    periodic lattice simulation. (b) maximum breather lattice point energy in parallel chains of atoms of propagation. (c) maximal atomic
    displacements of the breather solutions in the $x$-axis direction away
    from the equilibrium state. (d) maximal atomic displacements of
    the breather solutions in the $y$-axis direction away from the
    equilibrium state.}\label{fig:Eps}
\end{figure}

In the first study we fix parameter $\epsilon=0.05$ and vary the value
of $\gamma$, see Fig.\ \ref{fig:Eps}. In Figure \ref{fig:Eps}, we show
five numerical results for different values of gamma, i.e.\
$\gamma=0.4,0.5,0.6,0.7,0.8$. In Figure \ref{fig:EpsA} we plot the
number of sites the breather has passed versus time. Results show that
breather propagates faster as the value of $\gamma$ increases. In
Figures \ref{fig:EpsB}--\ref{fig:EpsD}, we plot the maximal lattice point energy and maximal displacements in the $x$ and $y$-axis
directions, indicated by the functions $\Delta{x}$ and $\Delta{y}$,
respectively, over the computational time of atoms on the atomic
chains $y_{m+k}$, where $k=0,1,2,3,4$. Due to symmetry considerations,
we have omitted from the figures the results in the $y_{m-k}$ chains and
the plots of the minimal displacement values in the $x$ and $y$-axis
directions.

Figure \ref{fig:EpsB} shows that most of the propagating breather lattice point energy is localized on the main chain $y_{m}$ where the breather
propagates and increases with the values of $\gamma$. Figures
\ref{fig:EpsC} and \ref{fig:EpsD} demonstrate large atomic
displacements of the breather solution on the main chain of atoms
$y_{m}$ in $x$-axis direction while there is almost zero displacement in the
$y$-axis direction. Equivalently to the energy results, Fig.~\ref{fig:EpsB} shows that displacements in the $x$-axis direction on the main
chain of atoms increase with larger values of $\gamma$. In contrast,
there is no strong evidence of increases in displacements in the
$y$-axis direction on the adjacent chain of atoms $y_{m+1}$ for larger
values of $\gamma$, see Fig.\ \ref{fig:EpsD}. Notice the scale
differences in Figs.\ \ref{fig:EpsC} and \ref{fig:EpsD}, and that
there are non vanishing displacements of atoms in chains $y_{m+2}$,
$y_{m+3}$ and $y_{m+4}$. This is due to the presence of low amplitude
phonons in the dynamics. Figure \ref{fig:Eps} strongly demonstrates
the quasi-one-dimensional nature of mobile breather solutions of the
lattice model (\ref{Hamilt}).

\begin{figure}
  \centering \subfigure[]{\label{fig:GamA}
    \includegraphics[trim=0.1cm 0.1cm 0.8cm 0.5cm,clip=true,width=0.315\textwidth]{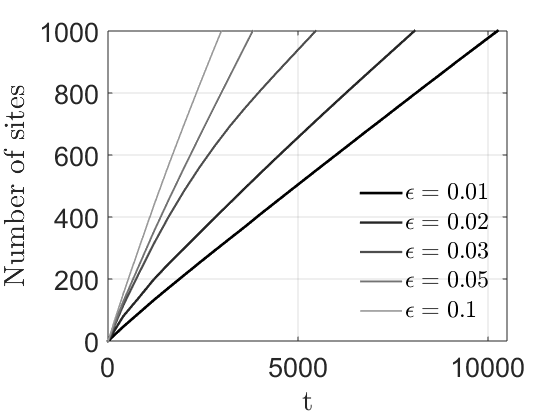}}
  \subfigure[]{\label{fig:GamB}
    \includegraphics[trim=0.1cm 0.1cm 0.8cm 0.5cm,clip=true,width=0.315\textwidth]{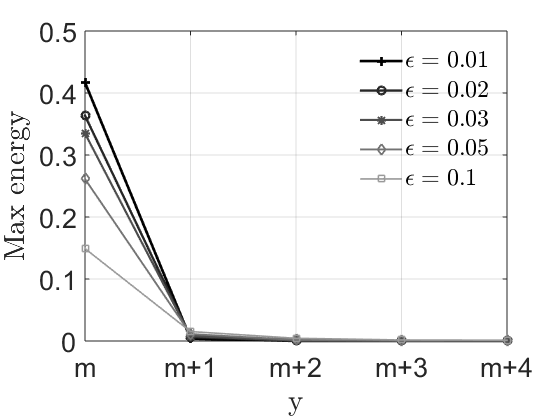}}
  \subfigure[]{\label{fig:GamC}
    \includegraphics[trim=0.1cm 0.1cm 0.8cm 0.5cm,clip=true,width=0.315\textwidth]{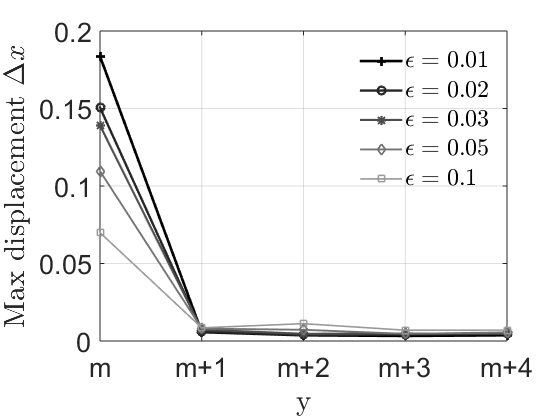}}
  \subfigure[]{\label{fig:GamD}
    \includegraphics[trim=0.1cm 0.1cm 0.8cm 0.5cm,clip=true,width=0.315\textwidth]{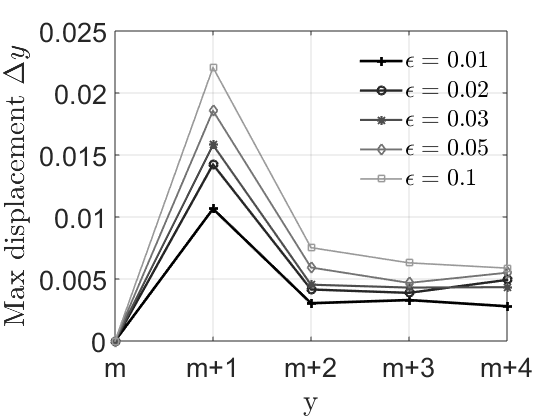}}
  \caption{Mobile breather properties for fixed excitation parameter
    value $\gamma=0.5$ and different values of the dimensionless parameter
    $\epsilon$. (a) breather travel distance versus time in periodic
    lattice simulation. (b) maximum breather lattice point energy in parallel chains
    of atoms of propagation. (c) maximal atomic displacements of the
    breather solutions in the $x$-axis direction away from the equilibrium
    state. (d) maximal atomic displacements of the breather solutions
    in the $y$-axis direction away from the equilibrium state.}\label{fig:Gam}
\end{figure}

We contrast the numerical results of Fig.\ \ref{fig:Eps} with
numerical results of Fig.\ \ref{fig:Gam}, where we have considered
numerical simulations of mobile breather solutions for excitation
parameter value $\gamma=0.5$ and various values of $\epsilon$. We
consider five values of $\epsilon$, that is, $\epsilon=0.01, 0.02, 0.03, 0.05, 0.1$. Figure \ref{fig:GamA} shows that the breather propagates faster with larger value of $\epsilon$, while the maximal energy on the main chain of atoms decreases with increasing value of $\epsilon$, see Fig.\ \ref{fig:GamB}. The same can be observed in Fig.\ \ref{fig:GamC} for atomic displacements in the $x$-axis direction. On the contrary, for larger values of $\epsilon$, displacements in the $y$-axis direction increases on adjacent chain of atoms $y_{m+1}$, as can be seen in Fig.\ \ref{fig:GamD}. This strongly indicates the increase of the 2D character of mobile breather
solutions in the dynamics with larger values of $\epsilon$. This transverse spreading is important when discussing breathers colliding on adjacent, parallel tracks. We confirm this observation in the study of breather energy scattering by breather-breather collisions below. In the present paper for conciseness, we consider only the values $\epsilon = 0.01, 0.05$.

\begin{figure}
\centering 
\subfigure[]{\label{fig:CollA} \includegraphics[width=0.48\textwidth]
   {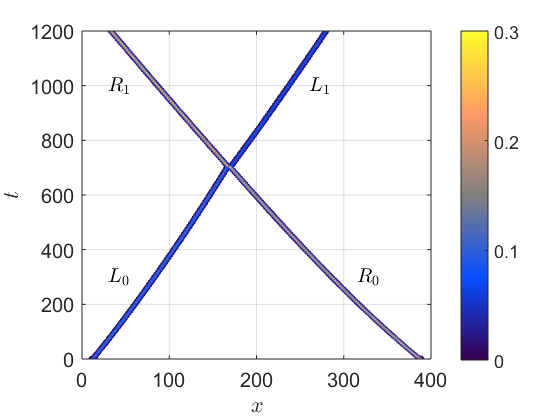}}
\subfigure[]{\label{fig:CollD} \includegraphics[width=0.48\textwidth]
   {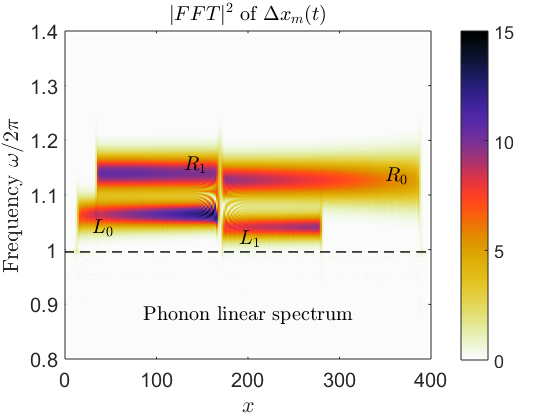}} 
 \caption{Simulation of two mobile breather head-to-head
   collision. (a) lattice point energy function on the main atomic
   chain of breather propagation. (b) frequency spectrum of atomic
   displacement function $\Delta x_m(t)$ in $x$-axis direction from
   equilibrium.  }\label{fig:Coll}
\end{figure}

The observations of mobile breather spectra and properties presented
in \cite{BaEiLe15} are demonstrated here with $\epsilon=0.05$ and a
head-to-head collision shown in Figure \ref{fig:Coll}. Consider a
periodic lattice of size $N_x=400$ and $N_y=32$ of atoms in their
equilibrium state and launch two atomic excitations (\ref{Pattern}) in
the middle of the lattice at each ends of chain $y_m$.  We indicate
left and right excitation parameter values by $\gamma_l$ and
$\gamma_r$, respectively of opposite signs, to set the breathers on a
collision course. We set $\gamma_{l}=0.4$ and $\gamma_{r}=-0.5$, and
integrate in time until $T_{end}=1200$. At around $t=700$, the breathers collide and reappear, see Figure
  \ref{fig:Coll}.  As in many soliton/breather collision scenarios, we
  cannot distinguish between the cases where the breathers pass
  through each other, or the case where one breather bounces off the
  other without passing through.  A more detailed study will be
  undertaken in future to determine if energy is exchanged or lost in
  this process.

In Figure \ref{fig:CollA} we plot the lattice point energy function in time of the atoms on the main chain $y_m$ where the
greatest energy of the breather solution is localized. We indicate by
$L_0$ and $R_0$ the left and right propagating breathers before
collision, respectively, while by $L_1$ and $R_1$ we indicate the same
breathers after the collision. Notice the slight change in the
breather propagation speed after the collision. The frequency spectrum is seen in Fig.~\ref{fig:CollD}, where we use the same notation to identify breather solutions. Note also the breather frequency focusing in time before and after the collision, as was observed in \cite{BaEiLe15}. In addition to breather atomic displacement spectrum we have identified the phonon linear spectrum band
of the dispersion relation calculated in \cite{BaEiLe15}.

The result of Figure \ref{fig:Coll} can be thought of as demonstrating
a strongly one-dimensional nature, despite the 2D nature of the
lattice.  Due to the chaotic nature of molecular dynamics, the
numerical observations, particularly at long times, are sensitive to
changes in the initial conditions and to round-off error.  This
motivates us to consider an ensemble of initial conditions as well as
different starting configurations, to study breather-breather collisions. 

For the ensemble, we draw two sets of normally distributed random
numbers $X,Y\sim\mathcal{N}(0,1)$ and scale them to normally
distributed numbers with mean values $\gamma_{l,r}$ and variance
$\sigma^2_{l,r}$. Thus we obtain two sets of the excitation parameter
value $\gamma\sim\mathcal{N}(\gamma_{l,r},\sigma_{l,r}^2)$ for
numerical simulations. For the following examples on the lattice
$N_x=200$ and $N_y=64$, we consider two sets of $2000$ random numbers
sampling the standard Cauchy distribution (i.e.~of the ratio $X/Y$)
and scale parameters equal to zero and one, respectively, with mean
values $\gamma_{l,r}=\pm0.5$ and variances $\sigma^{2}_{l,r}=0.002$.
Since a small amount of energy is present in the lattice, from phonons
generated from the initial excitations, we set atomic energy density
values to zero if the value is smaller than $0.01$. This value is
estimated from the numerical observations. Thus most of the phonon
energy is disregarded for the final energy averages and most of the
information comes from the breather solutions.

Using this set of initial velocities for {\em inline} collisions (as
in Fig.\ \ref{fig:Coll}) we did not observe scattering of breather
solutions into different crystallographic lattice directions despite a
visible spread of energy around the main chain of atoms of breather
propagation. Despite that, depending on the $\gamma_l$ and $\gamma_r$
values, we observed rich collision events such as the appearance of
just one moving or stationary breather, two breathers moving in the
same direction and one stationary breather together with one moving
breather. Predominantly, the most common case of all was the
appearance of two moving breathers after a collision.

\begin{figure}
\centering
\subfigure[]{\label{fig:EnCollC} \includegraphics[width=0.48\textwidth]
   {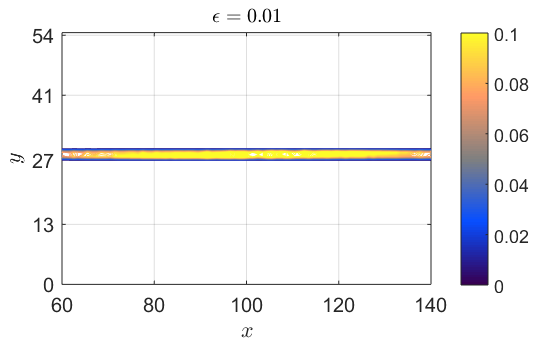}}
\subfigure[]{\label{fig:EnCollD} \includegraphics[width=0.48\textwidth]
   {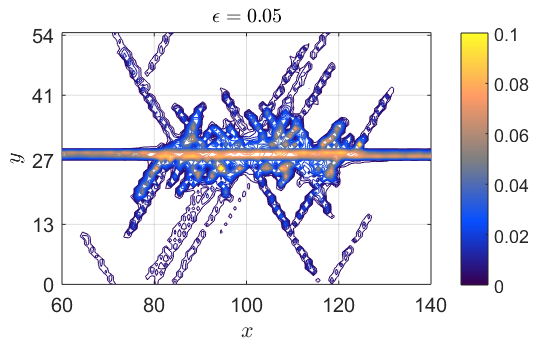}}  
\caption{Constrained ($H>0.01$) energy density function averaged over
   50 time snapshots and 2000 individual simulations of
   breather-breather collisions on adjacent chains of atoms. (a)
   simulation with $\epsilon=0.01$ until $T_{end}=1000$.  (b)
   simulation with $\epsilon=0.05$ until $T_{end}=500$.
 }\label{fig:EnColl}
\end{figure}

If instead we consider the scattering of two breathers on {\em
  adjacent} parallel lines, the results depend on the value of
$\epsilon$ used. For $\epsilon=0.01$, Fig.~\ref{fig:EnCollC}, we
observed no scattering of breather solutions into different
crystallographic lattice directions. However for $\epsilon=0.05$, we
observe breather scattering in all lattice directions,
Fig.~\ref{fig:EnCollD}.  Notice that dark energy lines arise {\em
  only} in these directions indicating propagating as well as
stationary breather solutions. Collisions may even lead to fast moving
breathers which due to periodic boundary conditions enter back into
the computational domain.

\begin{figure}
  \centering \subfigure[]{\label{fig:En1}
    \includegraphics[trim=0.5cm 0.cm 0.0cm 0.1cm,clip=true,width=0.34\textwidth]{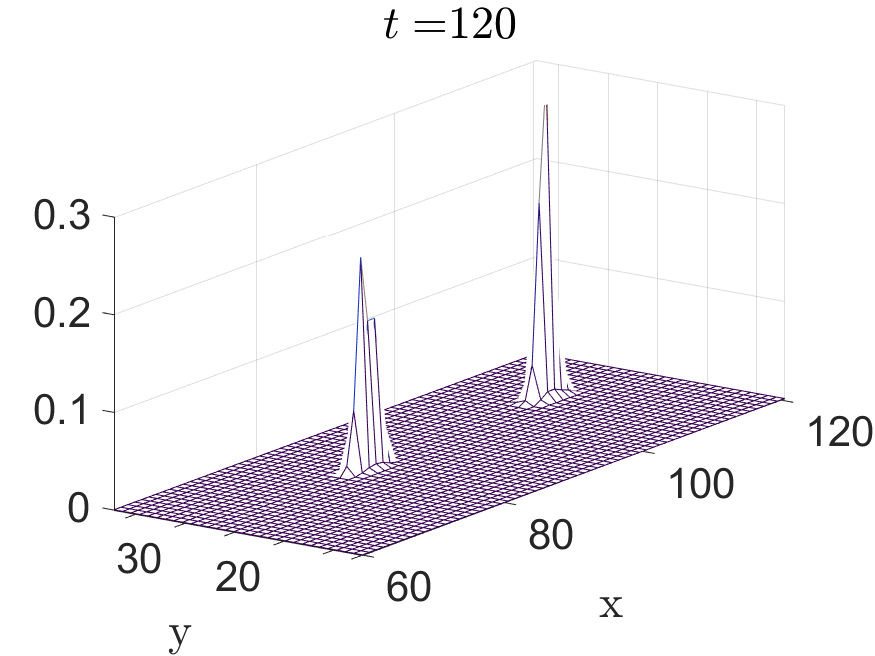}}
  \subfigure[]{\label{fig:En2}
    \includegraphics[trim=0.5cm 0.cm 0.0cm 0.1cm,clip=true,width=0.34\textwidth]{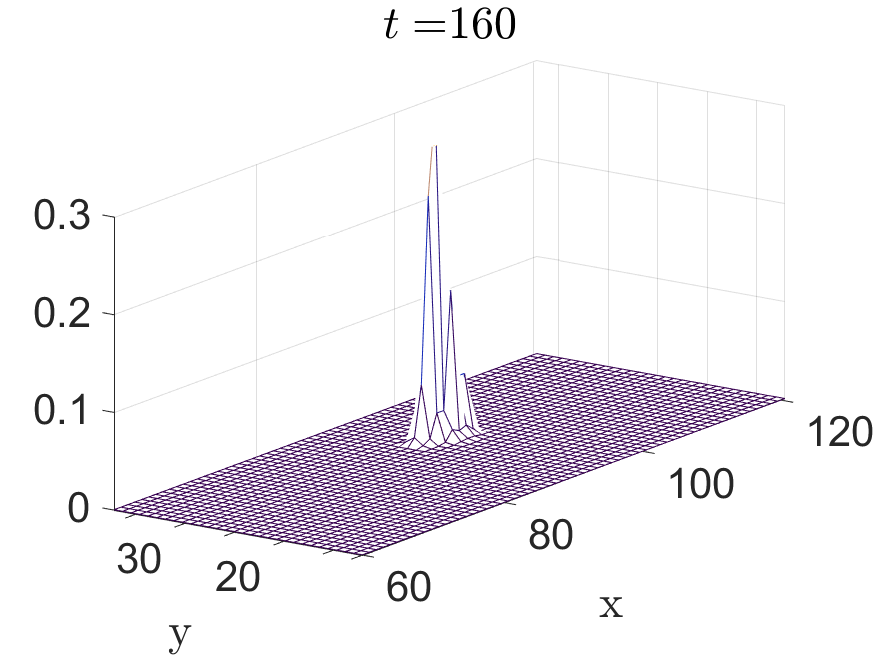}}
  \subfigure[]{\label{fig:En3}
    \includegraphics[trim=0.5cm 0.cm 0.0cm 0.1cm,clip=true,width=0.34\textwidth]{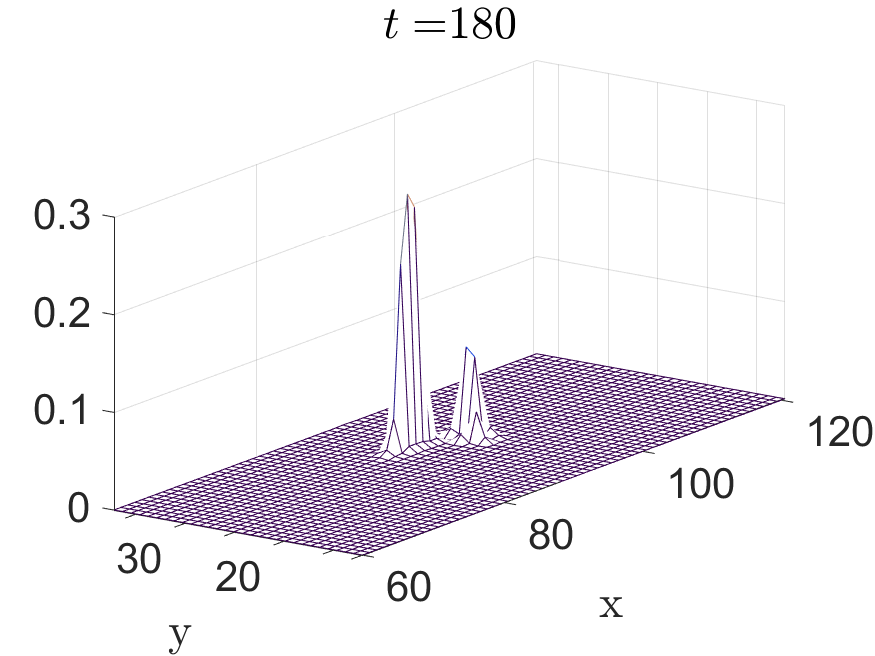}}
  \subfigure[]{\label{fig:En4}
    \includegraphics[trim=0.5cm 0.cm 0.0cm 0.1cm,clip=true,width=0.34\textwidth]{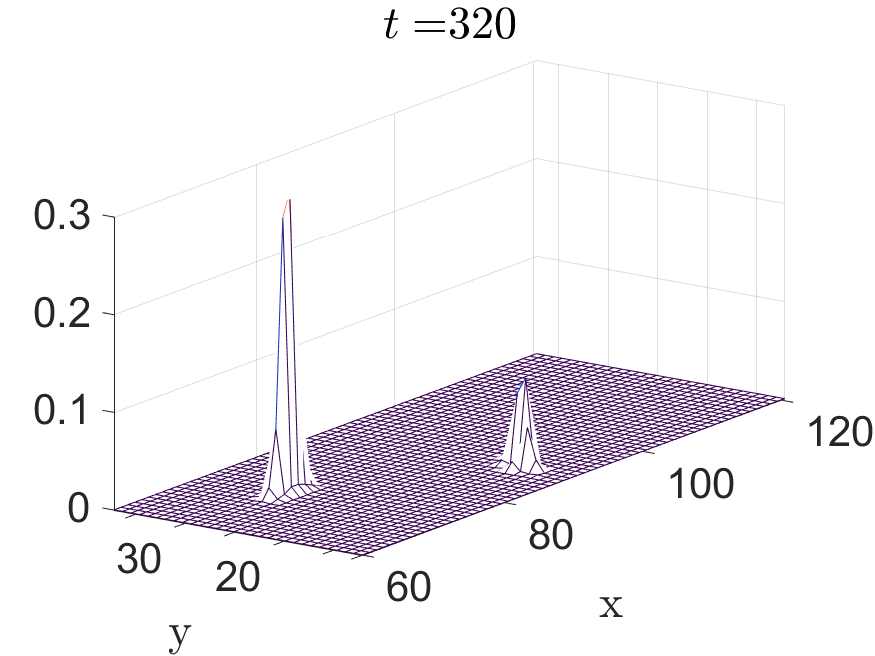}}
\caption{Lattice point energy density function of a single moving breather collision on adjacent parallel lines at four different times. (a) $t=120$. (b) $t=160$. (c) $t=180$. (d) $t=320$. Parameter values: $\epsilon=0.05$, $\gamma_l=0.475$ and $\gamma_r=-0.57$.}\label{fig:En}
\end{figure}

To better illustrate breather 2D scattering in Fig.~\ref{fig:EnCollD}
we consider a single collision with excitation parameter values:
$\gamma_l=0.475$ and $\gamma_r=-0.57$. In Figure \ref{fig:En} we plot
the energy density function at four different times. At time $t=120$,
Fig.~\ref{fig:En1}, two breathers are moving towards each
other. Around time $t=160$, Fig.~\ref{fig:En2}, both breathers
collide. The mobile breather from the right continues its path after
the collision, while the moving breather from the left gets scattered
with an angle, see Figures \ref{fig:En3} and \ref{fig:En4}.

These examples demonstrate the 2D properties of breather solutions, energy scattering by breathers and the importance of the parameter $\epsilon$. Figure
\ref{fig:EnCollD} confirms that the mobile breather's 2D character
increases for larger values of $\epsilon$, that is, for a stronger
interaction potential relative to a weaker on-site potential
energy. Since the Hamiltonian (\ref{Hamilt}) is time reversible, Fig.\
\ref{fig:EnCollD} also demonstrates breather-breather collisions with
an angle to each other when the time is reversed.

\begin{figure}
\centering 
\subfigure[]{\label{fig:SCollE001}\includegraphics[trim=1.8cm 2.5cm 1.5cm 2cm,clip=true,width=0.48\textwidth]
{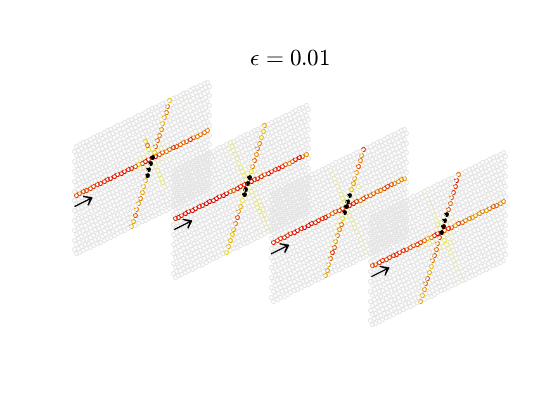}}
\subfigure[]{\label{fig:SCollE005}\includegraphics[trim=1.8cm 2.5cm 1.5cm 2cm,clip=true,width=0.48\textwidth]
{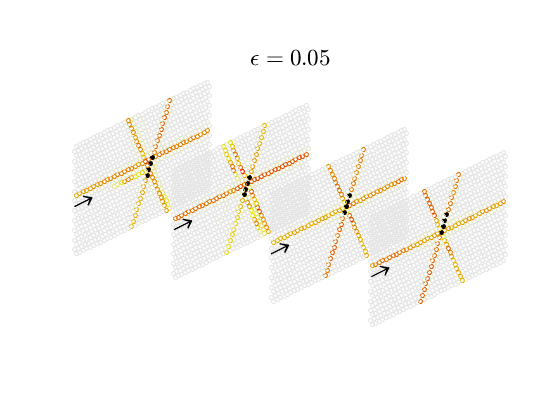}}
\caption{Constrained ($H>0.01$) energy density function averaged over
  50 time snapshots and 2000 individual simulations of mobile breather
  collisions with a stationary breather on a crystal axis at
  60$^\circ$ to the moving one. The stationary breather is at slightly
  different positions in each case. (a) simulation with
  $\epsilon=0.01$ until $T_{end}=1000$. (b) simulation with
  $\epsilon=0.05$ until $T_{end}=500$.}
\label{fig:SColl}
\end{figure}

To explore further breather scattering by breather-breather collisions in 2D
lattice model (\ref{Hamilt}) we consider simulations of mobile
breather collisions with a {\em stationary} breather at an angle to
the incoming in Fig.\ \ref{fig:SColl}. We consider four different
locations of the $(1,-2,2,-1)$ stationary excitation pattern indicated
by black circles while mobile breather propagation directions are
indicated by black arrows.

For this experiment we consider a smaller size lattice, $N_x=120$ and
$N_y=64$, and integrate until $T_{end}=500$, if $\epsilon=0.05$, and
until $T_{end}=1000$, if $\epsilon=0.01$.  We consider a mean
excitation value of $\gamma_{l}=0.5$ and $\gamma_{r}=-0.35$ where
$\gamma_r$ refers to the stationary breather and variances
$\sigma_{l}^2=0.001$ and $\sigma_{r}^2=0.00025$. We used a much
smaller variance value for the stationary breather compared to the
moving one to ensure that the excitation is not too large to generate
a mobile breather.

Because of the spread in incoming velocities/strengths etc., the
relative {\em phases} of the two breathers will also be different in
each simulation.

As above, we observe stronger breather scattering in the computations with $\epsilon=0.05$, Fig.~\ref{fig:SCollE005}, in contrast to the simulations with
$\epsilon=0.01$, Fig.~\ref{fig:SCollE001}, where the scattering is
predominately only in the lattice directions of both breathers.

Not only do we see scattering through a multiple of 60$^\circ$, but
the details of which track the breathers scatters to is very sensitive
to the velocity and phase of the incoming breather as well as the
position of the stationary breather. To better interpret the results
of Figure \ref{fig:SColl} we note that the energy plot illustrates a
rich variety of states, e.g., single stationary or moving breathers in
all lattice directions, two breathers moving with an angle to each
other and two breathers of which one is stationary.

Our study has given us a better understanding of particle-like tracks
in muscovite mica crystals. We demonstrate the importance of the
relative strengths of the interatomic force and of the force from the
surrounding atoms for the existence of long-lived propagating
breathers and their 2D collision properties.  Recent experimental work
by Russell et al.\ \cite{rra19} suggests strongly that breather-like
objects are important in real 3D crystals of several different
materials, displaying hyperconductivity and annealing effects at
finite temperatures despite a range of defects such as impurities,
dislocations, and crystal boundaries.  We plan to extend the current
model to one covering more realistic physical situations.

JB, during his postdoctoral research at the University of Edinburgh,
and BJL acknowledge the support of the Engineering and Physical
Sciences Research Council which has funded this work as part of the
Numerical Algorithms and Intelligent Software Centre under Grant
EP/G036136/1. JCE thanks Mike Russell for many useful conversations.
We are most grateful to an anonymous referee for many helpful comments
and corrections.

\bibliographystyle{unsrt}
\bibliography{BreatherColl_bibfile}

\end{document}